\documentclass[twocolumn,prl,preprintnumbers,floatfix]{revtex4-1}

\usepackage{graphicx}
\usepackage{dcolumn}
\usepackage{bm}
\usepackage{amsmath,amssymb,amsfonts}
\usepackage{epsf}

\newcommand{\vu}{{\mathbf u}}
\newcommand{\vv}{{\mathbf v}}
\newcommand{\vw}{{\mathbf w}}
\newcommand{\vx}{{\mathbf x}}

\newcommand{\vz}{{\mathbf z}}
\newcommand{\va}{{\mathbf a}}

\begin{document}

\title{Fluid Stretching as a L\'evy Process}

\author{Marco Dentz}
\affiliation{Spanish National Research Council (IDAEA-CSIC), 08034 Barcelona, Spain}
\email{marco.dentz@csic.es}
\author{Daniel R. Lester} 
\affiliation{School of Engineering,
  RMIT University, 3000 Melbourne, Victoria, Australia}
\author{Tanguy Le Borgne}
\affiliation{Geosciences Rennes, UMR 6118, Universit\'e de Rennes 1, CNRS, Rennes, France}
\author{Felipe P. J. de Barros}
\affiliation{Sonny Astani Department of Civil and Environmental
  Engineering, University of Southern California, 3620 S. Vermont
  Avenue, KAP 224B, Los Angeles, CA 90089, USA}

\date{\today}

\begin{abstract}
We study the relation between flow structure and fluid deformation in
steady two-dimensional random flows. Beyond the linear (shear flow)
and exponential (chaotic flow) elongation paradigms, we find a broad spectrum of stretching
behaviors, ranging from sub- to superlinear, which are dominated by
intermittent shear events. We analyze these behaviors from first
principles, which uncovers stretching as a result of the non-linear coupling between
Lagrangian shear deformation and velocity fluctuations along
streamlines. We derive explicit expressions for Lagrangian deformation
and demonstrate that stretching obeys a coupled continous time random
walk, which for broad distributions of flow velocities describes a
L\'evy walk for elongation. The derived model provides a direct link between the flow and deformation
statistics, and a natural way to quantify the impact of intermittent
shear events on the stretching behavior, which can have strong anomalous
diffusive character. 

\end{abstract}

\maketitle
The deformation dynamics 
and stretching history of material fluid elements are
fundamental for the understanding of hydrodynamic phenomena
ranging from scalar dispersion, pair
dispersion~\cite[][]{ShlesingerTurbulence1987, RastPinton2011, Thalabard:2014aa},
mixing~\cite[][]{Kleinfelter2005, Villermaux::Duplat::PRL::2006,
  Jha::PRL:::2011, deBarros::GRL::2012, LB2015, cirpkaprl2015} and
reaction~\cite[][]{Ranz79, Tartakovsky::2008, EngdahlPRE2014, Hidalgo2015} to the
alignment of material elements~\cite[][]{Lapeyre1999} and the distribution of stress in
complex fluids~\cite[][]{Truesdell:1992aa}.  
Fluid elements constitute the Lagrangian
support of a transported scalar. Thus, their deformation histories
determine the organization of the scalar distribution into lamellar
structures~\cite[][]{kalda00,
  Villermaux::Duplat::PRL::2003, Villermaux_12, LeBorgne2013}. Observed broad scalar
concentration distributions are a manifestation of a broad distribution of
stretching and compression rates and can explain intermittent patterns
of scalar increment distributions~\cite[][]{kalda00,
  Villermaux::Duplat::PRL::2003}. The temporal scaling of the average
elongation $\langle \ell(t) \rangle$ of material lines controls the decay of scalar variance, 
the effective kinetics of chemical reactions and the distribution of scalar gradients
\cite{Ottino::1989}. The mechanisms of
linear stretching due to persistent shear deformation, and exponential
stretching in chaotic flows have been well understood~\cite{Ottino::1989}. 
Observations of sub-exponential and non-linear fluid
elongation~\cite[][]{Duplat2000, Duplat::PF::2010,LeBorgne2013},
pair-dispersion~\cite[][]{ShlesingerTurbulence1987, Goto2004, RastPinton2011, Thalabard:2014aa,Afik2015}, 
and scalar variance decay~\cite[][]{nelkinPF1981, Vassilicos::2002}, however,
challenge these paradigms and ask for new dynamic frameworks. 
Even if stretching may be expected to be asymptotically exponential,
there generally exists a persistent pre-asymptotic algebraic mixing
regime ~\cite[][]{Vassilicos::2002}, which is critical as most mixing and associated chemical reactions are 
likely to occur at early times. 

While exponential stretching regimes are well understood, 
the theoretical description of algebraic stretching and mixing
behaviors is still debated and different mechanisms have been proposed
to describe it, including fractal/spiral mixing~\cite[e.g.][]{Vassilicos::2002}, 
non-sequential stretching~\cite[e.g.][]{Duplat::PF::2010}, and modified
Richardson laws~\cite[e.g.][]{nelkinPF1981}.
The dynamics of particle pair separation, for example, have been described using Levy processes and continuous time
random walks~\cite[][]{ShlesingerTurbulence1987, BofettaPRL2002,
  Thalabard:2014aa}.  Elongation time series for stretching in $d = 2$ dimensional
heterogeneous porous media flows have been modeled as geometric
Brownian motions~\cite{LB2015}.  

Most stochastic stretching models,
however, do not provide relations between the deformation dynamics and the local Lagrangian
and Eulerian deformations and flow structure. This means, the fluctuation
mechanisms that cause observed algebraic stretching are often not known. 
Broad velocity distributions as observed in disordered
media \cite{BG1990} and porous media flows \cite{Borgne::PRL::2008,
  Bijeljic::2011} lead to anomalous dispersion, which has been the subject of intense theoretical and
experimental studies~\cite[][]{BG1990, BESC97, Seymour::2004, Bijeljic::2011, 
deAnna2013}. The consequences for fluid stretching are much less known.  
Thus, we focus here on the relation between velocity fluctuations and fluid
deformation in non-helical steady random flows, such as steady $d = 2$
dimensional pore-scale and $d = 2$ and $d = 3$ dimensional Darcy-scale
flows in heterogeneous media~\cite[][]{Sposito1994, 3DJFM}. Such flows occur in natural
and engineered materials including porous and fractured
rocks~\cite[][]{Bear1972}, porous films,
carbon layers, chromatography, packed bed
reactors~\cite[][]{Brenner1997, Jakobsen2008}, biofilms and
biological tissue~\cite[][]{Vafai2010}. 
We derive a mechanism that leads to a broad range of
sub-exponential and power-law stretching behaviors. We formulate Lagrangian deformation in streamline
coordinates~\cite[][]{Winter:1982aa}, which relates elongation to
Lagrangian velocities and shear deformation. The consequences of this
coupling are studied in the framework of a continuous time
random walk (CTRW)~\cite[][]{MW1965, SL73.1, LevyReview2015} that
links transit times of material fluid elements to elongation through
Lagrangian velocities. We show that non-linear stretching behaviors
can be caused by broad velocity distributions. 

Our analysis starts with the equation of motion of a fluid particle in
a steady spatially varying flow field. The particle position
$\vx(t|\va)$ in the divergence-free flow field
$\vu(\vx)$ evolves according to the advection equation
\begin{align}
\label{advection:equation}
\frac{d \vx(t|\va)}{d t} = \vv(t), 
\end{align}
where $\vv(t) = \vu[\vx(t|\va)]$ denotes the Lagrangian
velocity. The initial condition is given by $\vx(t = 0|\va) =
\va$. The particle movement along a streamline can be formulated as 
\begin{align}
\label{st}
\frac{d s(t)}{d t} = v(t), && d t = \frac{ds}{v_s(s)},
\end{align}
where $s(t)$ is the distance travelled along the streamline, $v(t)
= |\vv(t)|$ and the streamwise velocity is $v_s(s) =
|\vv[t(s)]|$. With these preparations, we focus now on the evolution of the
elongation of an infinitesimal material fluid element, whose length
and orientation are described by the vector $\vz(t) = \vx(t|\va + \delta \va)
- \vx(t|\va)$. According to~\eqref{advection:equation}, its evolution
is governed by 
\begin{align}
\label{zeps}
\frac{d \vz(t)}{d t} = \boldsymbol \epsilon(t) \vz(t),
\end{align}
where $\boldsymbol \epsilon(t) = \nabla \vu[\vx(t|\va)]^\top = \nabla \vv(t)^\top$ is the
velocity gradient tensor. Note that $\mathbf z(t) = \mathbf
  F(t) \mathbf z(0)$ with $\mathbf F(t)$ the deformation tensor. Thus,
$\mathbf F(t)$ satisfies Eq.~\eqref{zeps} and the following
analysis is equally valid for the deformation tensor. The elongation $\ell(t)$ is given by
$\ell(t) = |\vz(t)|$. We transform the deformation process into the
streamline coordinate system~\cite[][]{Winter:1982aa}, which is attached to and rotates along the
streamline described by $\vx(t|\va)$,
\begin{align}
\vx^\prime(t) = \mathbf A^\top(t) \left[\vx(t) - \vx(t|\va)\right],
\end{align}
where the orthogonal matrix $\mathbf A(t)$ describes the rotation
operator which orients the $x_1$--coordinate with   
the orientation of velocity $\vv(t)$ along the streamline such that $\mathbf A(t) =
[\vv(t),\vw(t)]/v(t)$ with $\vw(t) \cdot \vv(t) =
0$ and $|\vw(t)| = v(t)$. 
From this, we obtain for $\vz^\prime(t) = \mathbf A^\top \vz(t)$ in
the streamline coordinate system
\begin{align}
\label{zp}
\frac{d \mathbf z^\prime(t)}{d t} = \left[ \mathbf Q(t)  + \tilde
  {\boldsymbol \epsilon}(t)\right] \mathbf z^\prime(t), 
\end{align}
where we defined $\tilde {\boldsymbol \epsilon}(t) = \mathbf A^\top(t)
\boldsymbol \epsilon(t) \mathbf A(t)$ and the antisymmetric tensor
$\mathbf Q(t) = \frac{d \mathbf A^\top(t)}{d t} \mathbf A(t)$. 
Thus, the velocity gradient tensor $\boldsymbol \epsilon(t)$
transforms into the streamline system as
$\boldsymbol \epsilon^\prime(t) = \mathbf Q(t)  + \tilde {\boldsymbol \epsilon}(t)$.
A quick calculation reveals that the components of $\mathbf Q(t)$ are
given by $Q_{12}(t) = - Q_{21}(t) = \tilde \epsilon_{21}(t)$, where we
use that $\frac{d \mathbf v(t)}{d t} = \boldsymbol \epsilon(t)
\mathbf v(t)$.  This gives for the velocity gradient in the
streamline system the upper triangular form 
\begin{align}
\boldsymbol \epsilon^\prime(t) =
\begin{bmatrix}
\tilde \epsilon_{11}(t) & \sigma(t)\\
0 & - \tilde \epsilon_{11}(t)
\end{bmatrix}, 
\label{eqn:2Dvelgrad}
\end{align}
where we define the shear rate $\sigma(t) =
\tilde \epsilon_{12}(t) + \tilde \epsilon_{21}(t)$ along the
streamline. 
Note that $\tilde \epsilon_{11}(t) = d v_s[s(t)] / d s$
by definition. Furthermore, due to the incompressibility of
$\vu(\vx)$, $\tilde \epsilon_{22}(t) = - \tilde \epsilon_{11}(t)$. 
For simplicity of notation, in the following we drop the
primes. The upper triangular form of $\boldsymbol \epsilon(t)$ as a direct result
of the transformation into the streamline system permits explicit
solution of~\eqref{zp} and reveals the dynamic origins of algebraic
stretching. 

\begin{figure}
\includegraphics[width = .42\textwidth]{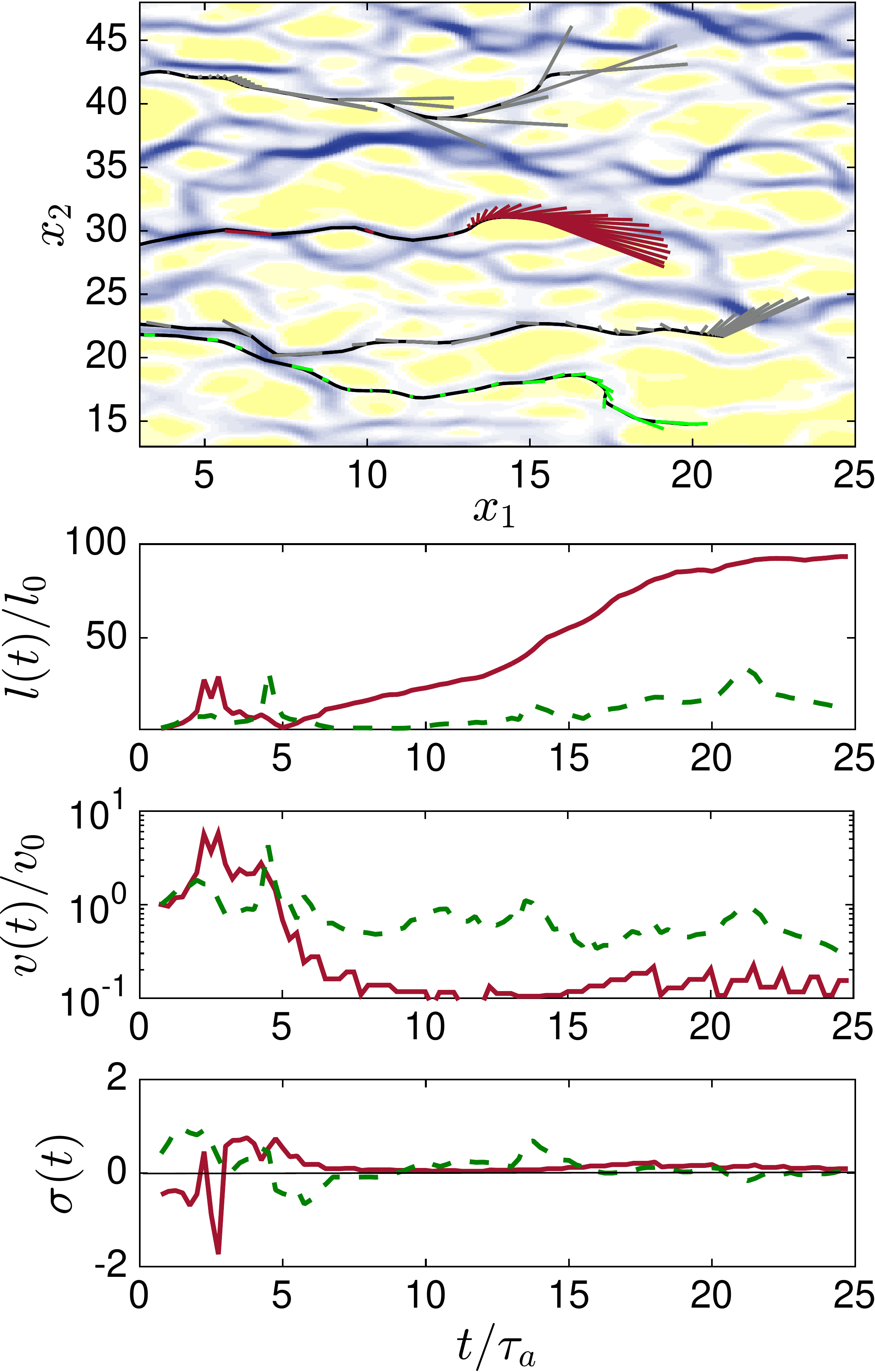}
\caption{(Left panel) Illustration of the evolution of length (rescaled)
  and orientation of
material strips along streamlines in a steady $d = 2$ dimensional
  divergence-free random flow~\cite[][]{LeBorgne2013}. The color
  scales illustrate the velocity magnitude decreasing from blue to
  yellow. Strips are drawn along streamlines at equidistant times. 
We observe persistent stretching in low velocity zones.  
This is reflected in the right panel, which illustrates (top) strip elongations
$\ell(t)$ for two distinct streamlines characterized by high (green
dashed) and low (red solid) velocities, (middle)
  strip velocity time series, and (bottom) shear deformation 
corresponding to the strip evolutions illustrated in the left panel by
the same colors. 
\label{Fig:illustration}}
\end{figure}

Thus, we can formulate the evolution equation~\eqref{zp}
of a material strip in streamline coordinates as
\begin{subequations}
\label{zsystem}
\begin{align}
d z_1(s) &= \frac{d v_s(s)}{v_s(s)} z_1(s) +
\frac{\sigma(s)}{v_s(s)} z_2(s) ds
\\
d z_2(s) & = - \frac{d v_s(s)}{v_s(s)} z_2(s), 
\end{align}
\end{subequations}
where we used~\eqref{st} to express $\vz(t) =
\vz[s(t)]$ in terms of the distance along the streamline. 
The system~\eqref{zsystem} can be integrated to 
\begin{subequations}
\label{prediction_Winter}
\begin{align}
\label{z1}
z_1(s) &= \frac{v_s(s)}{v_s(0)} \left[z_1(0) + z_2(0)
\int\limits_0^s d s^\prime \sigma(s^\prime)
\frac{v_s(0)^2}{v_s(s^\prime)^3} \right]
\\
z_2(s) &= \frac{v_s(0)}{v_s(s)}  z_2(0).
\label{z2}
\end{align}
\end{subequations}
Note that the deformation tensor $\mathbf F(t)$ in the streamline
system has also an upper triangular form. Its components can be
directly read off the system~\eqref{prediction_Winter}.
The angle of the strip $\vz(t)$ with respect to the streamline
orientation is denoted by $\phi(t)$ such that $z_1(t) = \ell(t)
\cos[\phi(t)]$ and $z_2(t) = \ell(t) \sin[\phi(t)]$. The initial strip
length and angle are denoted by $\ell_0$ and $\phi_0$. The strip length
is given by $\ell(t) \equiv \ell[s(t)]$ with $\ell(s) = [z_1(s)^2 +
z_2(s)^2]^{1/2}$.

The system~\eqref{prediction_Winter} is of general validity for $d = 2$ dimensional
steady flow fields. It reveals the mechanisms that lead to an increase
of the strip elongation, which is fully determined by the shear
deformation $\sigma(s)$ and the velocity $v_s(s)$ along the streamline. 
For a strip that is initially aligned with the streamline,
 $z_2(0) = 0$, the elongation is $\ell(s) = z_2(0) v_s(s)/v_s(0)$
 because $z_2(s) \equiv 0$ remains zero. This means $\ell(s)$ merely
 fluctuates without a net increase~\footnote[1]{The supplementary
   material gives some details on the calculation of elongation and
   the numerical random walk simulations.}. Only if the strip is oriented away from
the streamline can the streamwise velocity fluctuations be converted into
stretching. This identifies the integral term in~\eqref{z1} as the dominant contribution to the strip
elongation. It represents the interaction of shear
deformation and velocity with a linear contribution from the shear
rate and a non-linear contribution from velocity as
$1/v_s(s)^{3}$, which may be understood as follows. One power comes from the divergence
of streamlines in low velocity zones, which increases $z_2(s)$ and thus leads to
enhanced shear deformation. The second power is purely
kinematic due to the weighting with the residence time in a streamline
segment. The third power stems from the fact that shear deformation in 
low velocity segments is applied while the strip is compressed in
streamline direction. This deformation is then amplified as the strip is
stretched due to velocity increase. As a result of this non-linear coupling, the history of low
velocity episodes has a significant impact on the net stretching as quantified
by the integral term in~\eqref{z2}. This persistent effect is
superposed with the local velocity fluctuations. These mechanisms are illustrated in
Figure~\ref{Fig:illustration}. 
While for a stratified flow field with $\vu(\vx) = \vu(x_2)$ velocity
and shear deformation are constant along a streamline such that
$\ell(t) = [(z_1(0) + z_2(0) \sigma t)^2 +
z_2(0)^2]^{1/2}$, that is, it increases linearly with time, stretching
can in general be sub- or superlinear, depending on the duration of
low velocity episodes. In the following, we will analyze these behaviors in
order to identify and quantify the origins of algebraic stretching. 

To investigate the consequences of the non-linear coupling between
shear and velocity on the emergence of su-exponential stretching, we cast
the dynamics~\eqref{prediction_Winter} in the framework of a coupled
CTRW. Thus, we assume that the random flow field is stationary and
ergodic and consider fluid elements that move along ergodic
streamlines~\footnote[2]{For flows displaying open and closed
  streamlines such as the steady $d = 2$ dimensional Kraichnan model,
we focus on stretching in the subset of ergodic
streamlines. Stretching due to shear on closed streamlines is linear
in time.}.
We consider random flows $\vu(\vx)$ whose velocity fluctuations are controlled by a
characteristic length scale. We focus on the impact of 
broad velocity point distributions rather than on that of long range correlation \cite{BG1990, DentzBolster2010}.
This is particularly relevant for porous media flows. It has been observed
at the pore and Darcy scales that the streamwise velocity, that is, the velocity measured
equidistantly along a streamline follows a Markov process~\cite[][]{Borgne2008a,
  Borgne::PRL::2008,kangdentz11-prl, deAnna2013,
  Edery2014}.
Thus, if we choose a coarse-graining scale that is of the order of the
streamwise correlation length $\lambda_c$,~\eqref{st} can be discretized as 
\begin{align}
\label{ctrw}
s_{n+1} = s_n + \lambda_c, && t_{n+1} = t_n + \frac{\lambda_c}{v_n}.
\end{align}
The $v_n = v_s(s_n)$ are identical independently distributed random
velocities with the probability density function (PDF) $p_v(v)$. A
result of this spatial Markovianity is that the
particle movement follows a continuous time random walk
(CTRW)~\cite[][]{SL73.1,BESC97}. 
The PDF of streamwise velocities $p_v(v)$ is related to the
Eulerian velocity PDF $p_e(v)$ through flux weighting as $p_v(v) \propto
v p_e(v)$. The Eulerian velocity PDF in $d = 2$ dimensional pore-networks, for example, can be
approximated by a  Gaussian-shaped distribution, which breaks down for small
velocities~\cite[][]{Araujo2006}. For Darcy scale porous and fractured media the velocity PDF can be
 characterized by algebraic behaviors at small
 velocities~\cite[][]{BESC97, Edery2014, KangPRE2015}, which
 implies a broad distribution of transition times $\tau_n = \lambda_c
 / v_n$. Note, however, that the proposed CTRW stretching mechanisms
 is of general nature and valid for any velocity distribution
 $p_v(v)$.  
Thus, in order to extract the deformation dynamics, we coarse-grain
the elongation process along the streamline on the correlation scale
$\lambda_c$. This gives for the strip coordinates~\eqref{prediction_Winter}
%
\begin{subequations}
\label{zn}
\begin{align}
z_1(s_n) &= z_1(0) \frac{v_n}{v_0} + z_2(0)
\frac{v_n v_0}{v_c^2} \sigma_c \tau_v r_n
\label{z1n}
\\
z_2(s_n) &= z_2(0) \frac{v_0}{v_n},
\label{z2n}
\end{align}
\end{subequations}
with $v_c$ and $\sigma_c$ a characteristic velocity and shear rate, and
$\tau_v = \lambda_c / v_c$ a characteristic advection time. The process $r_n$, which
results from the integral term in~\eqref{z1}, describes the coupled CTRW  
\begin{align}
\label{LevyWalk}
r_{n+1} = r_n + \frac{v_c^3}{\sigma_c} \frac{\sigma_{n}}{v_n^3}, &&
t_{n + 1} = t_n + \frac{\lambda_c}{v_n}.  
\end{align}
The elongation at time $t$ is given by $\ell(t) = [z_1(s_{n_t})^2 + z_2(s_{n_t})^2]^{1/2}$.
It is observed over several $2 d$ flows that the shear rate may be
related to the streamwise velocity as 
$\sigma_n = \xi_n \sigma_c (v_n/v_c)^{\hat \alpha}$ with $\hat
\alpha \approx 1$, $\sigma_c$ a characteristic shear rate, and $\xi_n$ an identical
independent random variable that is equal to $\pm 1$ with equal
probability. The average shear rate $\langle \sigma_n
\rangle = 0$ due to the stationarity of the random flow field
$\vu(\vx)$. Thus,~\eqref{LevyWalk} denotes a coupled CTRW whose increments
$\rho_n \equiv r_{n+1} - r_n$ are related to the transition times $\tau_n =
\lambda_c / v_n$ as
\begin{align}
\label{dr}
\rho_n = \xi_n \left({\tau_n}/{\tau_v}\right)^{\alpha}, && \alpha = 3
- \hat \alpha.  
\end{align}
It has the average $\langle \rho_n
\rangle = 0$ and absolute value $|\rho_n| = (\tau_n /
\tau_v)^\alpha$. The joint PDF of the elongation increments $\rho$ and
transition times $\tau$ is then given by 
\begin{align}
\psi(\rho,\tau) = \frac{1}{2} \delta\left[|\rho| - (\tau / \tau_v)^\alpha\right] \psi(\tau), 
\end{align}
where $\delta(\rho)$ denotes the Dirac delta distribution. 
The transition time PDF $\psi(\tau)$ is related to the streamwise
velocity PDF $p_v(v)$ as $\psi(\tau) = \lambda_c \tau^{-2} p_v(\lambda_c /
\tau)$. 

In the following, we consider a streamwise velocity PDF that behaves
as $p_v(v) \propto (v/v_c)^{\beta -1}$ for $v$ smaller than the
characteristic velocity $v_c$. Such a power-law is a model for the low
end of the velocity spectra in disordered media \cite{BG1990} and porous media flows
\cite{BCDS2006, Bijeljic::2011, Edery2014}. 
Note however that the derived CTRW-based deformation mechanism is valid for any velocity
distribution.  
The relation between the streamwise and
Eulerian velocity PDFs, $p_v(v) \propto v p_e(v)$ implies that $\beta
\geq 1$ because $p_e(v)$ needs to be integrable in $v = 0$. 
The corresponding transition time PDF $\psi(\tau)$
 behaves as $\psi(\tau) \propto (\tau/\tau_v)^{-1-\beta}$ for
$\tau > \tau_v = \lambda_c/v$ and decreases sharply for $\tau <
\tau_v$. Due to the constraint $\beta > 1$, the
mean transition time $\langle \tau \rangle < \infty$ is always finite,
which is a consequence of fluid mass conservation. For transport in
highly heterogeneous pore Darcy-scale porous media
values for $\beta$ between $0$ and $2$ have been
reported~\cite[][]{BCDS2006, Bijeljic::2011}. It has been found that decreasing medium
heterogeneity leads to a sharpening of the  transition time PDF and
increase of the exponent $\beta$~\cite[][]{Edery2014} with $\beta >
1$. With these definitions, the coupled CTRW~\eqref{LevyWalk} describes a Levy
walk. 

\begin{figure}
\includegraphics[width = .49\textwidth]{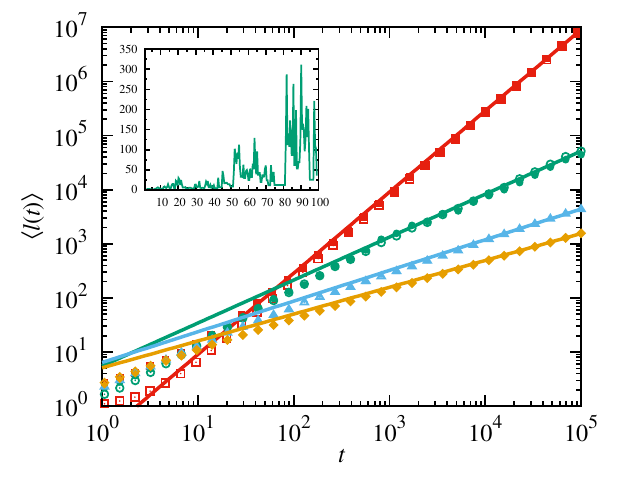}
\caption{Evolution of the (open symbols) mean elongation $\langle
  \ell(t) \rangle = \langle [z_1(s_{n_t})^2 + z_2(s_{n_t})^2]^{1/2}\rangle$ with $\ell_n$ given
  obtained from numerical Monte-Carlo simulations using~\eqref{zn}
  and~\eqref{LevyWalk} for a uniform distribution of initial strip orientations $\phi \in
  [-\pi/2,\pi/2]$ and a Gamma PDF of streamwise velocity. The full symbols are obtained from the 
  approximation~\eqref{ellapprox} for (squares) $\beta = 3/2$, (circles)
$\beta = 5/2$, (triangles) $\beta = 7/2$ and (rhombi) $\beta =
9/2$. The solid lines indicate the late time power-law behaviors of
$\langle \ell(t) \rangle \propto t^{3 - \beta}$ for $1 < \beta < 2$,
$\langle \ell(t) \rangle \propto t^{2/\beta}$ for $2 < \beta < 4$, and
$\langle \ell(t) \rangle \propto t^{1/2}$ for $\beta > 4$. The inset
illustrates the evolution of $\ell(t)$ in a single realization of the
velocity field $v_n$ for $\beta = 5/2$ for an initial orientation of
$\phi = 0$. \label{Fig:elongation}}
\end{figure}

Figure~\ref{Fig:elongation} shows the evolution of the
average elongation $\langle \ell(t) \rangle$ for $\alpha = 0$ and different
values of $\beta$ obtained from numerical Monte-Carlo simulation
using~\eqref{zn} and the Levy walk~\eqref{LevyWalk} for the evolution
of the strip coordinates based on a Gamma PDF of streamwise
velocities~\footnote[1]{}. The mean
elongation shows a power-law behavior and
increases as $\langle \ell(t) \rangle \propto t^\nu$. 
As discussed above, long episodes of small velocity maintain the strip
in a favorable shear angle, which leads to a strong stretching. These
dynamics are quantified by the L\'evy walk process~\eqref{LevyWalk},
which relates strong elongations to long transition times, i.e., 
small streamwise velocities, through~\eqref{dr}.
This is also illustrated in the inset of Figure~\ref{Fig:elongation},
which shows the elongation of a single material strip. The elongation
events increase with increasing time as a consequence of the
coupling~\eqref{dr} between stretching and transition time. This is an
intrinsic property of a CTRW characterized by a broad $\psi(\tau)$; the
transition times increase as time increases, and thus, through the Levy
walk coupling also the stretching increments.  
In fact, the strip length can be approximated by~\footnote[1]{}
\begin{align}
\label{ellapprox}
\ell(t) \approx \ell_0 + \frac{\sigma_c  \tau_v^2 \langle v
  \rangle}{\langle \tau \rangle v_c} |z_2(0)||r_{n_t}|.  
\end{align}
%
The leading behavior of the mean elongation $\langle \ell(t) \rangle$ of a material element
is directly related to the mean absolute moment of $r(t)$ as
$\langle \ell(t) \rangle \propto \langle |r_{n_t}| \rangle$. Thus,
even though $r_{n_t}$ is in average $0$, the addition of large
elongation events in its absolute value $|r(t)|$, which correspond to episodes of low
velocities, leads in average to an algebraic increase of $\ell(t)$ as
detailed in the following. 

The statistics of the Levy walk~\eqref{LevyWalk} have been analyzed in
detail in Ref.~\cite[][]{DLBLdB:PRE2015} for $\alpha > 0$ and
$\beta > 0$. Here, $\beta$ is restricted to $\beta > 1$ due to fluid
mass conservation. Furthermore, we consider $\alpha \geq 1$.  The
scaling of the mean absolute moments of $r_{n_t}$
depends on the $\alpha$ and $\beta$ regimes. 

If the exponent $\beta > 2 \alpha$, which means a relatively weak
heterogeneity,  we speak of a weak coupling between the elongation
increment $\rho_n$ and the transition time $\tau_n$ in~\eqref{dr}. In
this case, the strip elongation behaves as
$\langle \ell(t) \rangle \propto t^{1/2}$. 
We term this behavior here diffusive or normal stretching. For $\alpha
= 2$ as employed in the numerical simulations this means that $\beta
> 4$. The coupled
Levy-walk~\eqref{LevyWalk} reduces essentially to a Brownian motion
because the variability of transition times is low so that the
coupling does not lead to strong elongation events. Note that scalar dispersion in
this $\beta$--range is normal~\cite[][]{SL73.1, BCDS2006}. 

For strong coupling, this means $\beta < 2 \alpha$ and thus stronger flow
heterogeneity, it has been shown~\cite[][]{DLBLdB:PRE2015} that the density of
$r_{n_t}$ is characterized by two scaling forms, one that
characterizes the bulk behavior and a different one for large
$r_{n_t}$. As a consequence, we need to distinguish the cases of
$\beta$ larger and smaller than $\alpha$. Also, the scaling of
$|r_{n_t}|$ cannot be obtained by dimensional analysis.  In fact,
$r_{n_t}$ has a strong anomalous diffusive
character~\cite[][]{DLBLdB:PRE2015}. 

For $\alpha < \beta < 2 \alpha$ the scaling
behavior of the mean elongation is
$\langle \ell(t) \rangle \propto t^{\alpha/ \beta}$. 
This means for $\alpha = 2$, the stretching exponent $\nu$ is between
$1/2$ and $1$, the $\beta$--range is $2 < \beta < 4$. It interesting
to note that scalar dispersion in this range is normal as well. Here,
the frequency of low velocity regions is high enough to increase stretching
above the weakly coupled case, but not to cause super-diffusive scalar
dispersion.  

For $1 < \beta < \alpha$ in contrast, the mean elongation scales as
~\cite[][]{DLBLdB:PRE2015} 
$\langle \ell(t) \rangle \propto t^{1 + \alpha - \beta}$. 
The stretching exponent is between $1$ and $\alpha$, this means
stretching is stronger than for shear flow. The range of scaling
exponents $\nu$ of the mean elongation
here is $1/2 \leq \nu < \alpha$. Specifically, $\alpha
\approx 2$ implies that stretching is
super-linear for $1 < \beta < 2$, this means faster than by a pure
shear flow, for which $\nu = 1$. Here the presence of low velocities
in the flow leads to enhanced stretching and at the same time to
superdiffusive scalar dispersion.

In summary, we have presented a fundamental mechanism for
power-law stretching in random flows through intermittent shear events, which may explain 
algebraic mixing processes observed across a range of heterogeneous flows. 
We have shown that the non-linear coupling between streamwise
velocities and shear deformation implies that stretching follows
L\'evy walk dynamics, which explains
observed algebraic stretching behaviors that can range from diffusive
to super-diffusive scalings, $\langle \ell(t)
\rangle \propto t^{\nu}$ with $1/2 \leq \nu < 2$.  
The derived coupled stretching CTRW can be parameterized in terms of
the Eulerian velocity and deformation
statistics and provides a link between anomalous dispersion and fluid
deformation. The presented analysis
demonstrates that the dynamics of
fluid stretching in heterogeneous flow fields is much richer than the  
paradigmatic linear and exponential behaviors. 
The non-linear coupling between deformation and shear,
The fundamental mechanism of intermittent shear events,
which is at the root of non-exponential stretching, is likely present in
a broader class of fluid flows. 

\begin{acknowledgments}
MD acknowledges the support of the European Research Council (ERC)
through the project MHetScale (contract number 617511). TLB acknowledges Agence
National de Recherche (ANR) funding through the project
ANR-14-CE04-0003-01. 
\end{acknowledgments}
%
\end{document}



\title{Supplementary Material: Fluid Stretching as a L\'evy Process}
\author{Marco Dentz}
\affiliation{Spanish National Research Council (IDAEA-CSIC), 08034 Barcelona, Spain}
\email{marco.dentz@csic.es}
\author{Daniel R. Lester} 
\affiliation{School of Civil, Environmental and Chemical Engineering,
  RMIT University, 3000 Melbourne, Victoria, Australia}
\author{Tanguy Le Borgne}
\affiliation{Geosciences Rennes, UMR 6118, Universit\'e de Rennes 1, CNRS, Rennes, France}
\author{Felipe P. J. de Barros}
\affiliation{Sonny Astani Department of Civil and Environmental
  Engineering, University of Southern California, 3620 S. Vermont
  Avenue, KAP 224B, Los Angeles, CA 90089, USA}

\date{\today}

\begin{abstract}
This supplementary material gives details on the derivation of the
equations for the strip deformation in the streamline coordinate
system, the approximations for the calculation of the average strip
elongation, and the numerical random walk particle tracking simulations. 
\end{abstract}

\maketitle

\section{Deformation in the Streamline Coordinate System}
Here we briefly present the key steps leading to the explicit
solutions for the evolution of a material strip in the streamline
coordinate system. First, we note that the deformation rate tensor
$\boldsymbol \epsilon'(t)$ in the streamline coordinate system is
triangular. This can be seen as follows. As derived in the main text,
$\boldsymbol \epsilon'(t)$  is given by 
\begin{align}
\boldsymbol \epsilon'(t) = \mathbf Q(t) + \tilde \epsilon(t),
\end{align}
where $\mathbf Q(t) = d \mathbf A^\top(t)/dt \mathbf A(t)$ and $\tilde
\epsilon(t) = \mathbf A^\top(t)
\boldsymbol \epsilon(t) \mathbf A(t)$. Note that
$\mathbf A(t) = [\mathbf e_v(t),\mathbf e_w]$, with
$\mathbf e_v(t) = \vv(t)/v(t)$ and $\mathbf e_w(t) =
\vw(t)/v(t)$ and $\vv(t) \cdot \vw(t) = 0$. Thus, we obtain directly that $Q_{11}(t) = Q_{22}(t) = 0$ and
$Q_{12}(t) = -Q_{21}(t) = \mathbf e_w(t) d \mathbf e_v(t)/dt$, where 
\begin{align}
\frac{d \mathbf e_v(t)}{dt} &= -\frac{d \ln v(t)}{d t} \mathbf e_v(t) +
  \frac{1}{v(t)} \frac{d \vv(t)}{dt} = -\frac{d \ln v(t)}{d t} \mathbf e_v(t) +
  \frac{1}{v(t)} \boldsymbol \epsilon(t) \vv(t) 
\\
&= -\frac{d \ln v(t)}{d
  t} \mathbf e_v(t) 
+ \boldsymbol \epsilon(t) \mathbf e_v(t)
\end{align}
From the latter we obtain directly $Q_{12}(t) = \tilde
\epsilon_{21}(t)$, which gives the triangluar form of $\boldsymbol \epsilon'(t)$. 

The equations for the strip components $z_1'(t)$ and $z_2'(t)$ then
are given by 
\begin{align}
\frac{d z_1'(t)}{d t} & = \epsilon'_{11}(t) z_1'(t) + \sigma(t) z_2'(t)
\\
\frac{d z_2'(t)}{d t} & = - \epsilon'_{11}(t) z_2'(t). 
\end{align}
We transform now $d t = ds/v_s(s)$ and note that $\epsilon_{11}' =
dv_s/ds$ in order to obtain
\begin{align}
\label{z1}
\frac{d z_1'(s)}{d s} & = v_s(s)^{-1}\frac{d v_s(s)}{ds} z_1'(s) + \frac{\sigma(s)}{v_s(s)} z_2'(s)
\\
\frac{d z_2'(s)}{d s} & = - v_s(s)^{-1} \frac{d v_s(s)}{ds} z_2'(s). 
\end{align}
This system can be integrated. We start with the second equation for $z_2'(s)$, which can be written as 
\begin{align}
\frac{d z_2'(s)}{z_2'(s)}  = - \frac{d v_s(s)}{v_s(s)}. 
\end{align}
Thus, it can be directly integrated to 
\begin{align}
\ln[z_2'(s) / z_2'(0)] =  - \ln[v_s(s) / v_s(0)]. 
\end{align}
Taking the natural exponential on both sides gives 
\begin{align}
z_2'(s) =  \frac{v_s(0)}{v_s(s)}z_2'(0).
\end{align}
Inserting the latter into~\eqref{z1}, we obtain  
\begin{align}
\label{z1p}
\frac{d z_1'(s)}{d s} & = v_s(s)^{-1}\frac{d v_s(s)}{ds} z_1'(s) + \frac{\sigma(s) v_s(0)}{v_s(s)^2} z_2'(0). 
\end{align}
This equation can be solved by separation of variables. Thus, we write $z_1'(s) = f(s) g(s)$, where $f(s)$ 
satifies 
\begin{align}
\frac{d f(s)}{d s} & = v_s(s)^{-1}\frac{d v_s(s)}{ds} f(s).
\end{align}
Its solution is 
\begin{align}
f(s) = f(0) \frac{v_s(s)}{v_s(0)}. 
\end{align}
By inserting $z_1'(s) = f(0) v_s(s)/v_s(0) g(s)$ into~\eqref{z1p} gives for $g(s)$ the equation 
\begin{align}
\label{z1pp}
\frac{d g(s)}{d s} = \frac{1}{f(s)} \frac{\sigma(s) v_s(0)}{v_s(s)^2} z_2'(0) =  f(0)^{-1} \frac{\sigma(s) v_s(0)^2}{v_s(s)^3} z_2'(0)
\end{align}
Integration of the latter yields
\begin{align}
\label{z1pp2}
g(s) = g(0) +  f(0)^{-1} z_2'(0) \int\limits_0^s ds' \frac{\sigma(s') v_s(0)^2}{v_s(s')^3} 
\end{align}
and thus for $z_1'(s)$
\begin{align}
\label{z1ppp}
z_1'(s) = g(0) f(0) \frac{v_s(s)}{v_s(0)} +   z_2'(0) \frac{v_s(s)}{v_s(0)} \int\limits_0^s ds' \frac{\sigma(s') v_s(0)^2}{v_s(s')^3} 
\end{align}
The integration constants $g(0)f(0)$ are determined by the initial condition $z_1'(0)$ such that 
\begin{align}
\label{z1pppp}
z_1'(s) = z_1'(0) \frac{v_s(s)}{v_s(0)} +   z_2'(0)
  \frac{v_s(s)}{v_s(0)} \int\limits_0^s ds' \frac{\sigma(s')
  v_s(0)^2}{v_s(s')^3} .
\end{align}
%

\section{Strip Elongation}

In order to derive Eq. (14) in the main text for the strip elongation, we use the fact
that $v_{n_t} / v_0$ in~(10a) in the main text evolves in average towards
$1$, and that $v_0/v_{n_t}$ in~(10b) in the main text evolves in average slower than
$|r_{n_t}|$ as detailed below. Thus, we disregard $v_{n_t} / v_0$ and $v_0/v_{n_t}$ as
subleading. Furthermore, as shown below, we use that $v_s(s) v_s(0)$ 
converges in average towards the constant $ \langle v \rangle
\lambda_c/\langle \tau \rangle $ in order to obtain expression (14) in
the main text. 

\subsection{Average Elongation Perpendicular to the Streamline}

We first determine the average deformation of the material strip
perpendicular to the streamline, $\langle |z_2(t)| \rangle$. It is given
by 
\begin{align}
\langle |z_2(t)| \rangle = \langle |z_2(0)| \rangle \left\langle \frac{v_0}{v_{n_t}}
\right\rangle =  \langle z_2(0) \rangle
\sum\limits_{n = 0}^\infty \left\langle \frac{v_0}{v_{n}}
  \int\limits_0^t d t^\prime \delta(t^\prime - t_n) \mathbb I(0
  \leq t - t^\prime < \tau_n)
\right\rangle. 
\end{align}
It reads in Laplace space as
\begin{align}
\frac{\langle |z_2(\lambda)| \rangle}{\langle |z_2(0)| \rangle} &= 
\sum\limits_{n = 0}^\infty \left\langle \frac{v_0}{v_{n}} \exp(-
  \lambda t_n) \frac{1 - \exp(-\lambda \tau_n)}{\lambda}
\right\rangle
\\
& = 
\frac{1 - \psi(\lambda)}{\lambda} +  \langle v_0 \exp(-\lambda \tau_0) \rangle
\sum\limits_{n = 1}^\infty \psi(\lambda)^{n-1} \left\langle
  \frac{1}{v_{n}} \frac{1 - \exp(-\lambda \tau_n)}{\lambda}
\right\rangle
\\
& = 
\frac{1 - \psi(\lambda)}{\lambda} +  \frac{\langle \tau_0^{-1}
  \exp(-\lambda \tau_0) \rangle}{1 - \psi(\lambda)}
\left\langle
  \tau \frac{1 - \exp(-\lambda \tau)}{\lambda}
\right\rangle
\\
& = 
\frac{1 - \psi(\lambda)}{\lambda} +
\frac{1}{\lambda}\frac{\langle \tau_0^{-1}
  \exp(-\lambda \tau_0) \rangle}{1 - \psi(\lambda)}
\left[\langle \tau \rangle + \frac{d \psi(\lambda)}{d \lambda} \right]
\end{align}
If $\langle \tau^2 \rangle < \infty$, we obtain asymptotically 
\begin{align}
\langle |z_2(\lambda)| \rangle 
 \approx 
\frac{\langle |z_2(0)| \rangle \langle v_0 \rangle \langle \tau^2
  \rangle}{\lambda s_0 \langle \tau \rangle}. 
\end{align}
Thus, the long time value is given by 
\begin{align}
\lim_{t \to \infty}\langle |z_2(t)| \rangle 
 \approx 
\frac{\langle |z_2(0)| \rangle \langle v_0 \rangle \langle \tau^2
  \rangle}{s_0 \langle \tau \rangle}. 
\end{align}

In this Letter, we employ the Gamma distribution of streamwise
velocities
\begin{align}
p_v(v) = \frac{(v/v_c)^{\beta - 1} \exp(-v/v_c)}{v_c \Gamma(\beta)}
\end{align}
The velocity moments are given by 
\begin{align}
\label{mv}
\langle v^n \rangle = \frac{v_c^n \Gamma(\beta+n)}{\Gamma(\beta)}.  
\end{align}
The moments of the transition times $\tau = s_0 / v$ are given by 
\begin{align}
\label{mtau}
\langle \tau^n \rangle = \frac{\tau_v^n \Gamma(\beta-n)}{\Gamma(\beta)}.  
\end{align}
for $\beta > 2$. 

Thus, for the gamma velocity PDF employed in this Letter, we obtain
for the long time value 
\begin{align}
\label{z2approx}
\lim_{t \to \infty} \langle |z_2(t)| \rangle 
 = \frac{\langle |z_2(0)| \rangle \Gamma(\beta + 1) \Gamma(\beta -
   2)}{\Gamma(\beta) \Gamma(\beta - 1)}. 
\end{align}

For $1 < \beta < 2$, the second moment $\langle \tau^2 \rangle$ is not
finite. In this case, we obtain in the limit $\lambda \tau_v \ll 1$
for the Laplace transform of $\psi(\tau)$ 
\begin{align}
\psi(\lambda) = 1 - \langle \tau \rangle \lambda + a_\beta (\lambda
\tau_v)^\beta. 
\end{align}
Thus, in the limit of $\lambda \tau_v \ll 1$ and $1 < \beta < 2$ we
obtain for $\langle |z_2(\lambda)| \rangle$ in leading order  
\begin{align}
\langle |z_2(\lambda)| \rangle 
 \approx \frac{\langle |z_2(0)| \rangle \langle v_0 \rangle}{s_0\lambda^2 \langle \tau
   \rangle} \left[\langle \tau
   \rangle + \frac{d \psi(\lambda)}{d \lambda} \right] \propto
 \lambda^{\beta - 3}, 
\end{align}
which gives in real time the behavior
\begin{align}
\langle |z_2(t)| \rangle \propto t^{2 - \beta}.  
\end{align}
%

\begin{figure}
\includegraphics[width = .45\textwidth]{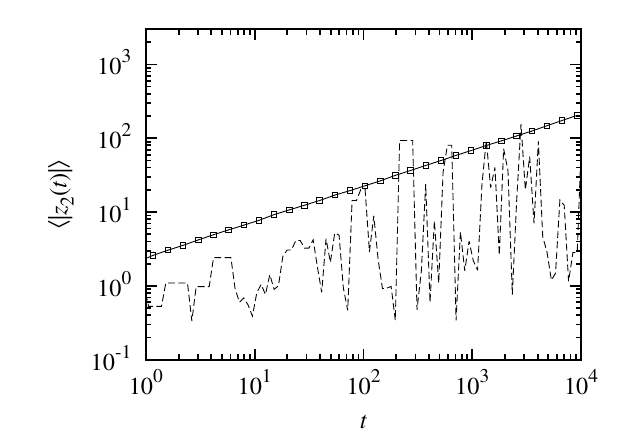}
\includegraphics[width = .45\textwidth]{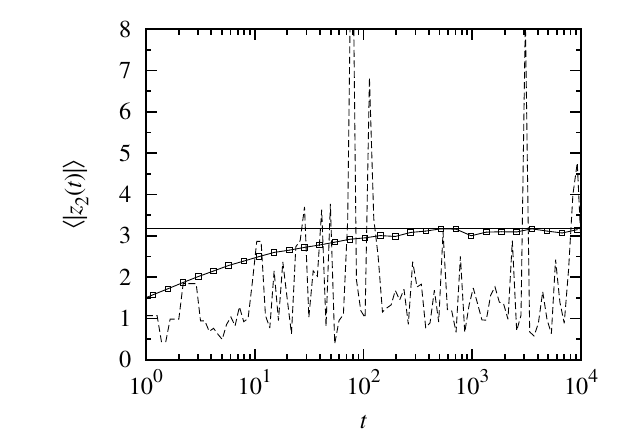}
\label{Fig:z2}
\caption{(Symbols) Average elongation perpendicular to the streamline for (left
  panel) $\beta = 3/2$ and (right panel) $\beta = 5/2$. (Dashed)
  Evolution of elongation in a single realization for $z_1(0) = 0$ and
$z_2(0) = 1$. The horizontal solid line in the right panel indicate the asymptotic
value~\eqref{z2approx}.}
\end{figure}

\subsection{Average Elongation Along the Streamline}

Now we consider the contribution $\delta z_1(t) = v_{n_t} /
v_0$, which quantifies the elongation for a material strip that is
initially aligned with the streamline, that is $z_2(0) = 0$. 
Similarly as for $\langle z_2(t) \rangle$, we obtain for the
Laplace transform of $\langle \delta z_1(t) \rangle$ 
\begin{align}
\frac{\langle |\delta z_1(\lambda)| \rangle}{\langle |z_1(0)| \rangle} &= 
\frac{1 - \psi(\lambda)}{\lambda} +  \frac{\langle \tau_0
  \exp(-\lambda \tau_0) \rangle}{1 - \psi(\lambda)}
\left\langle
  \frac{1 - \exp(-\lambda \tau)}{\lambda \tau }
\right\rangle
\\
& = 
\frac{1 - \psi(\lambda)}{\lambda} + \frac{1}{\lambda} \frac{\langle \tau_0
  \exp(-\lambda \tau_0) \rangle}{1 - \psi(\lambda)}\left[
\frac{\langle v \rangle}{s_0} - \left\langle \tau^{-1} \exp(-\lambda \tau)
\right\rangle \right]
\\
& = 
\frac{1 - \psi(\lambda)}{\lambda} - \frac{1}{\lambda [1 - \psi(\lambda)]}
\frac{d \psi_0(\lambda)}{d \lambda}\left[
\frac{\langle v \rangle}{s_0} - \left\langle \tau^{-1} \exp(-\lambda \tau)
\right\rangle \right]
\end{align}
In the limit of $\lambda \ll 1$ and $1 < \beta < 2$ we obtain in
leading order 
\begin{align}
\langle \delta z_1(\lambda) \rangle 
& \approx \frac{\langle |z_1(0)| \rangle}{\lambda}. 
\end{align}
Thus, we obtain asymptotically 
\begin{align}
\label{dz1approx}
\lim\limits_{t \to \infty} \langle \delta z_1(t) \rangle 
& = \langle |z_1(0)| \rangle. 
\end{align}
%

\begin{figure}
\includegraphics[width = .45\textwidth]{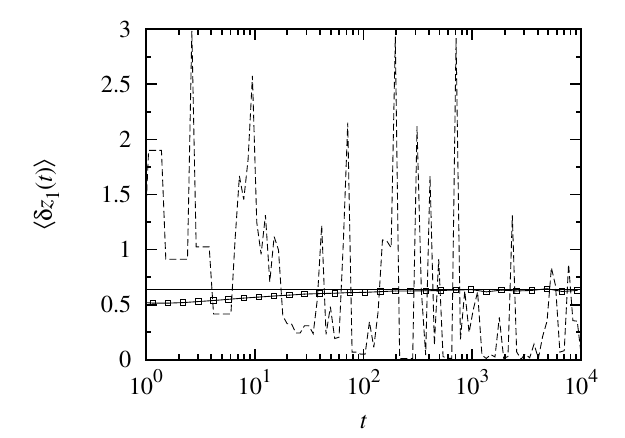}
\includegraphics[width = .45\textwidth]{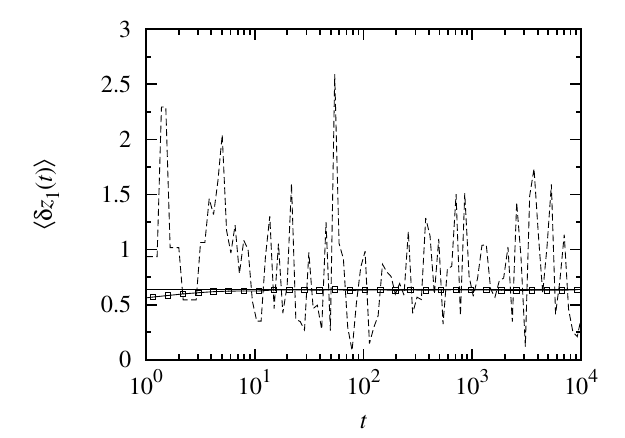}
\label{Fig:dz1}
\caption{(Symbols) Average elongation along the streamline for (left
  panel) $\beta = 3/2$ and (right panel) $\beta = 5/2$. (Dashed)
  Evolution of elongation in a single realization for $z_1(0) = 1$ and
$z_2(0) = 0$. The horizontal solid lines indicates the asymptotic
values~\eqref{dz1approx}. }
\end{figure}

\subsection{Average Velocity Cross-Product Along the Streamline}

Here, we determine the average velocity cross-product $C_v(t) = \langle v(0)
v(t) \rangle$, which can be written as
\begin{align}
C_v(t) = \langle v_0 v_{n_t} \rangle = \sum\limits_{n = 0}^\infty \left\langle
  v_0 v_n
  \int\limits_0^t \delta(t^\prime - t_n) \mathbb I(0 \leq t - t^\prime
  < \tau_n) \right\rangle 
\end{align}
Its Laplace transform is given by 
\begin{align}
C_v(\lambda) &= \sum\limits_{n = 0}^\infty \left\langle
  v_0 v_n \exp(- \lambda t_n) \frac{1 - \exp(-\lambda \tau_n)}{\lambda} \right\rangle 
\\
& = 
\frac{1 - \psi(\lambda)}{\lambda} +  \langle v_0 \exp(-\lambda \tau_0) \rangle
\sum\limits_{n = 1}^\infty \psi(\lambda)^{n-1} \left\langle
  v_{n} \frac{1 - \exp(-\lambda \tau_n)}{\lambda}
\right\rangle
\\
& = 
\frac{1 - \psi(\lambda)}{\lambda} +  s_0^2 \frac{\langle \tau_0^{-1}
  \exp(-\lambda \tau_0) \rangle}{1 - \psi(\lambda)}
\left\langle
  \tau^{-1} \frac{1 - \exp(-\lambda \tau)}{\lambda}
\right\rangle
\\
& = 
\frac{1 - \psi(\lambda)}{\lambda} +
\frac{s_0^2}{\lambda}\frac{\langle \tau_0^{-1}
  \exp(-\lambda \tau_0) \rangle}{1 - \psi(\lambda)}
\left[\frac{\langle v \rangle}{s_0} - \left\langle \tau^{-1} \exp(-\lambda \tau)
\right\rangle \right]
\end{align}
In the limit of $\lambda \ll 1$, we obtain
\begin{align}
C_v(\lambda) \approx \frac{1}{\lambda} \langle v_0 \rangle
\frac{s_0}{\langle \tau \rangle}. 
\end{align}
Thus, the asymptotic long time value is given by 
\begin{align}
\lim_{t \to \infty} C_v(t) = 
\frac{s_0 \langle v_0 \rangle }{\langle \tau \rangle}. 
\end{align}
For the gamma velocity PDF employed in this Letter, we obtain
from~\eqref{mv} and~\eqref{mtau} 
\begin{align}
\label{v0vnapprox}
\lim_{t \to \infty} \langle v(0) v(t) \rangle = v_c^2
\frac{\Gamma(\beta + 1)}{\Gamma(\beta - 1)}.  
\end{align}
%

\begin{figure}
\includegraphics[width = .45\textwidth]{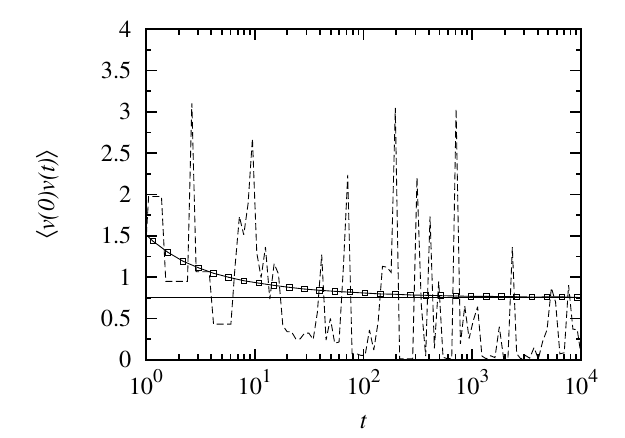}
\includegraphics[width = .45\textwidth]{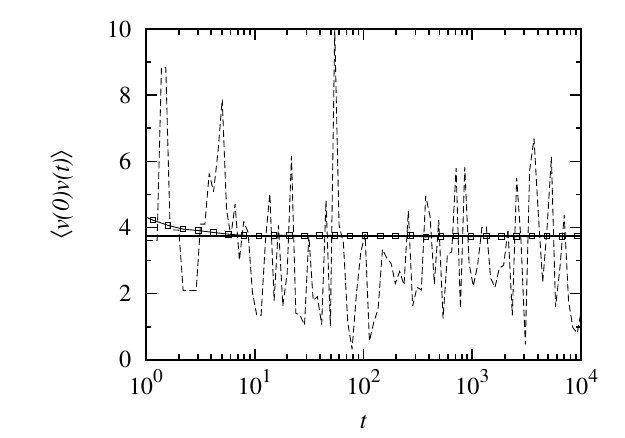}
\label{Fig:R1}
\caption{(Symbols) Average velocity cross-product for (left
  panel) $\beta = 3/2$ and (right panel) $\beta = 5/2$. (Dashed)
  Evolution of the cross-product in a single realization for $z_1(0) = 0$ and
$z_2(0) = 1$. The horizontal solid lines indicate the asymptotic
values~\eqref{v0vnapprox}.}
\end{figure}

\section{Numerical Simulations}
The numerical Monte-Carlo simulations illustrating the average strip
elongation in Figure 2 model the streamwise
velocity PDF by the Gamma distribution
\begin{align}
p_v(v) = \frac{(v/v_c)^{\beta - 1} \exp(-v/v_c)}{v_c\Gamma(\beta)},
\end{align}
which yields for the transition time distribution
\begin{align}
\psi(\tau) = \frac{1}{\tau_v \Gamma(\beta)}
\frac{\exp(-\tau_v/\tau)}{(\tau/\tau_v)^{1+\beta}}. 
\end{align}
We set $v_c = 1$ and $s_0 = 1$ such that $\tau_v = 1$. The initial
strip orientation angle $\phi$ is uniformly distributed in
$[-\pi/2,\pi/2]$. 
